\documentclass[a4paper,11pt]{article}
\pdfoutput=1 

\usepackage{jinstpub} 

\usepackage{lineno}
\usepackage{url}

\title{Caribou - A versatile data acquisition system for silicon pixel detector prototyping}

\author[a, 1]{Y. Otarid,\note{Corresponding author.}}
\author[b]{M. Benoit,}
\author[c]{E. Buschmann,}
\author[c]{H. Chen,}
\author[a]{D. Dannheim,}
\author[e]{T. Koffas,}
\author[e]{\mbox{R. St-Jean,}}
\author[d]{S. Spannagel,}
\author[c]{S. Tang,}
\author[d]{T. Vanat}

\affiliation[a]{CERN,\\Esplanade des Particules, Geneva ,Switzerland}
\affiliation[b]{ORNL,\\1 Bethel Valley Rd, Oak Ridge, United States}
\affiliation[c]{BNL,\\98 Rochester St, Upton, United States}
\affiliation[d]{DESY,\\Notkestrasse 85, Hamburg, Germany}
\affiliation[e]{Carleton University,\\1125 Colonel By Dr, Ottawa, Canada}

\emailAdd{younes.otarid@cern.ch}

\abstract{Caribou is a versatile data acquisition system used in multiple collaborative frameworks (CERN EP R\&D, DRD3, AIDAinnova, Tangerine) for laboratory and test-beam qualification of novel silicon pixel detector prototypes. The system is built around a common hardware, firmware and software stack shared accross different projects, thereby drastically reducing the development effort and cost. It consists of a custom Control and Readout (CaR) board and a commercial Xilinx Zynq System-on-Chip (SoC) platform. The SoC platform runs a full Yocto distribution integrating the custom software framework (Peary) and a custom FPGA firmware built within a common firmware infrastructure (Boreal). The CaR board provides a hardware environment featuring various services such as powering, slow-control, and high-speed data links for the target detector prototype. Boreal and Peary, in turn, offer firmware and software architectures that enable seamless integration of control and readout for new devices. While the first version of the system used a SoC platform based on the ZC706 evaluation board, migration to a Zynq UltraScale+ architecture is progressing towards the support of the ZCU102 board and the ultimate objective of integrating the SoC functionality directly into the CaR board, eliminating the need for separate evaluation boards. This paper describes the Caribou system, focusing on the latest project developments and showcasing progress and future plans across its hardware, firmware, and software components.}


\proceeding{The Eleventh International Workshop On Semiconductor Pixel Detectors for Particles and Imaging\\
  Nov 18-22, 2024\\
  Strasbourg, France}

\begin{document}
\maketitle
\flushbottom

\section{Introduction}
\label{sec:introduction}
Caribou is a modular and versatile data acquisition (DAQ) system designed for prototyping silicon pixel detectors \cite{Benoit, Fiergolski, Vanat, Buschmann}. Built around the Control and Readout (CaR) board, it provides essential interfaces and power supplies for a wide range of detector devices. At its core, Caribou features a System-on-Chip (SoC) running embedded Linux, combined with a software application and user-specific FPGA firmware. This integration of programmable logic and high-level software control streamlines detector testing and debugging by offering configurable firmware blocks, comprehensive software interfaces, and reusable hardware components.\\
Originally developed for the ATLAS Inner Tracker (ITk) and CLIC vertex detector projects, Caribou’s open-source architecture now supports various detector R\&D applications. By minimizing the effort required for the base system setup, it allows researchers to focus on device integration and testing. The system is widely used within the DRD3 collaboration on semiconductor detectors for laboratory and high-rate beam tests, facilitating seamless integration of new silicon-pixel-detector prototypes.

\section{System Architecture}
\label{sec:caribou_system_architecture}

\begin{figure}[h]
	\centering
	\includegraphics[width=0.8\textwidth]{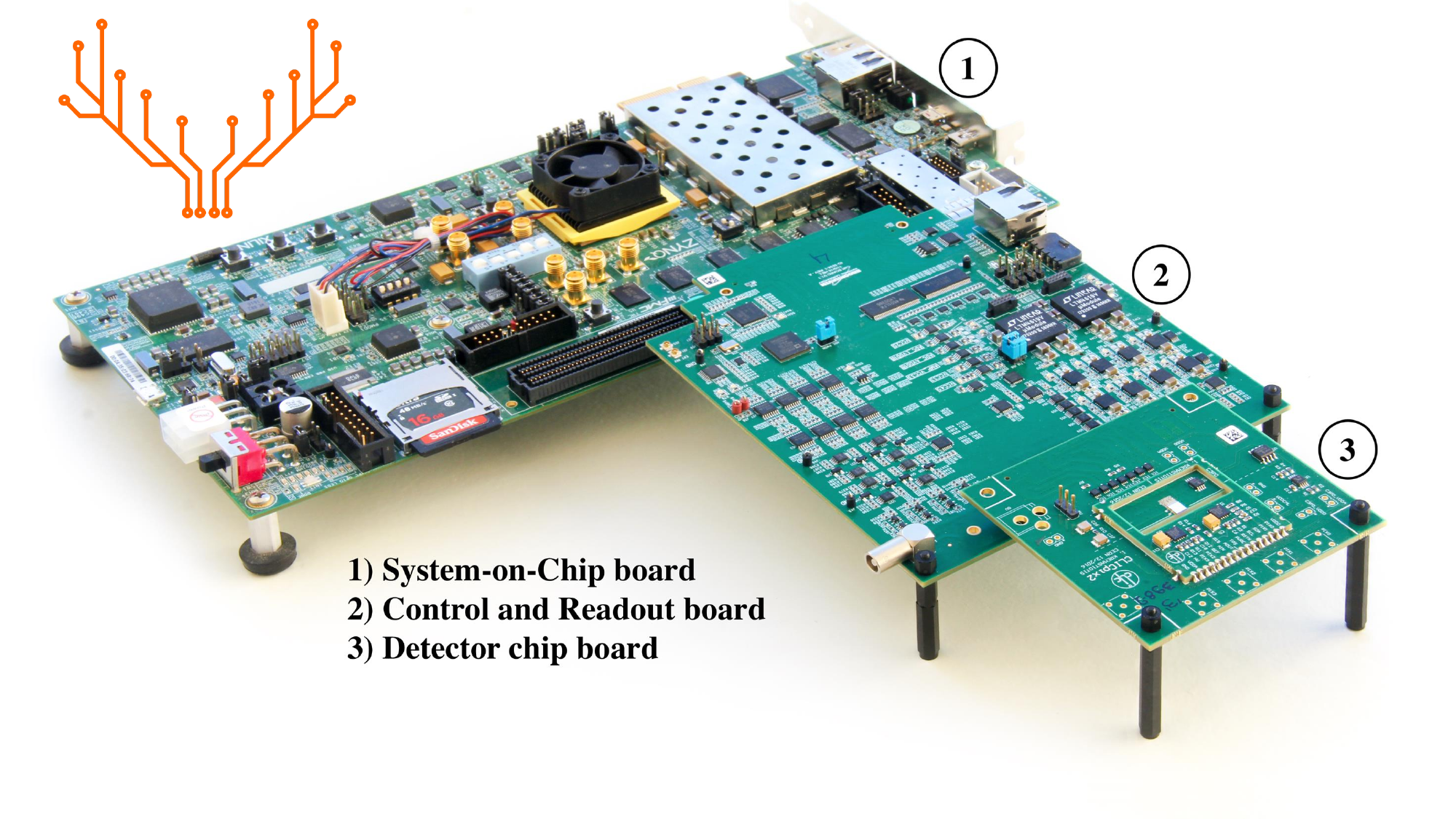} 
	\caption{Hardware components of the Caribou DAQ system}
	\label{fig:caribou_system}
\end{figure}

\begin{figure}[h]
	\centering
	\includegraphics[width=0.8\textwidth]{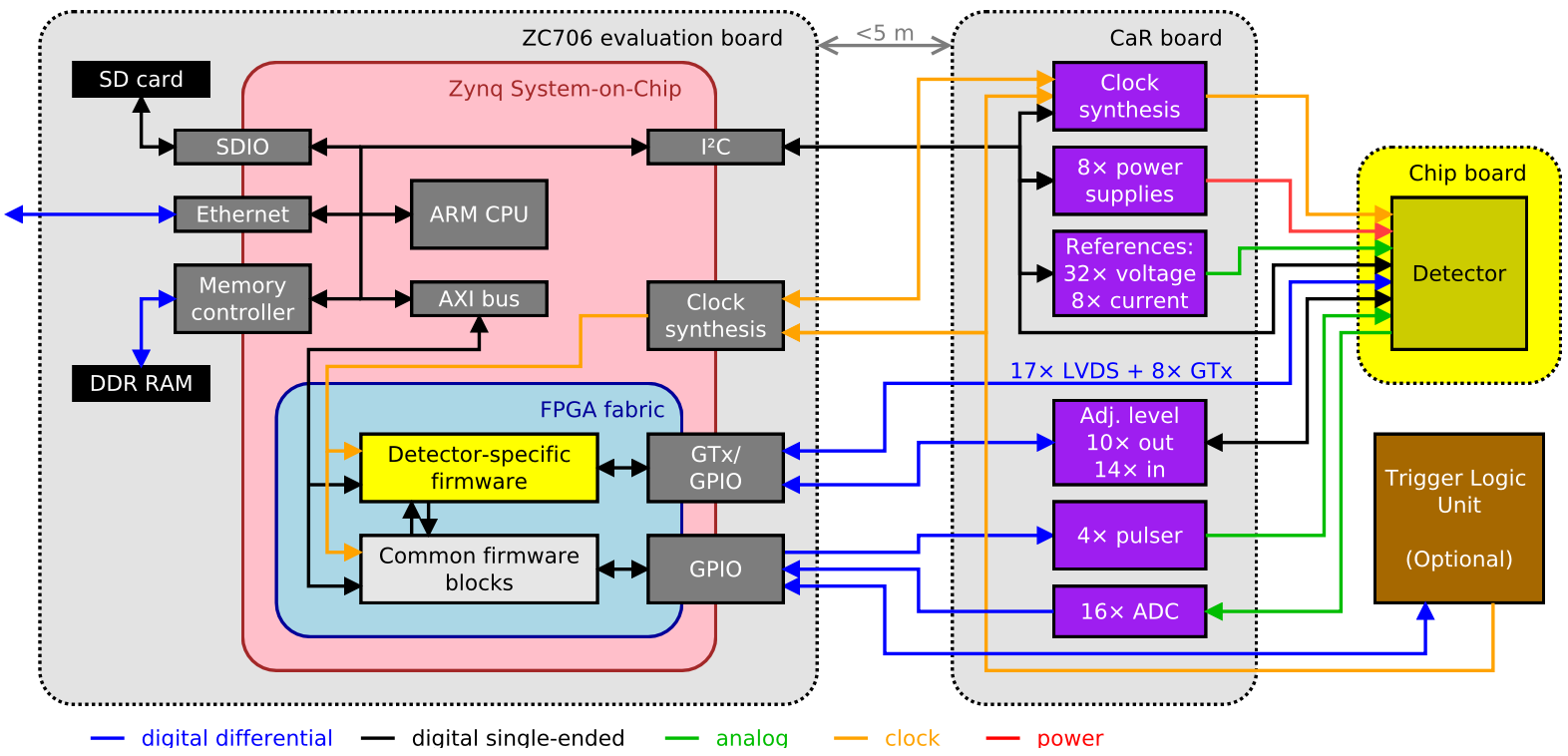} 
	\caption{Simplified block diagram of the Caribou hardware architecture \cite{Dannheim}}
	\label{fig:caribou_system_architecture}
\end{figure}

The Caribou system architecture is designed for flexibility and seamless integration with various silicon pixel detector prototypes. The CaR board interfaces with the SoC evaluation board via an FMC connector, with an optional cable extension for custom configurations. It offers essential components such as programmable power supplies, high-speed ADCs, injection pulsers, and a comprehensive set of voltage and current references. By integrating these critical functions into a single board, Caribou reduces the need for additional external electronics, simplifying detector integration. In addition to the SoC connection, the CaR board links to a chip board via a SEARAY connector. This low-cost, user-designed PCB hosts the detector prototype and incorporates all necessary passive components for its control and readout. Figure \ref{fig:caribou_system} and \ref{fig:caribou_system_architecture} show a picture of the Caribou system setup and a simplified overview diagram of its hardware architecture, respectively. As of today, more than 15 detector prototypes in more than 14 institutes have been characterized using the Caribou system.

A key strength of Caribou lies in its software stack, which pairs the embedded Linux OS with the Peary DAQ framework \cite{Peary Gitlab}. Peary’s Hardware Abstraction Layer (HAL) simplifies interaction with hardware peripherals by representing them as objects in C++. This abstraction enables rapid development of device-specific functions while maintaining a robust structure for managing multiple devices, logging, and interfacing through both CLI and API. Furthermore, the system's FPGA fabric supports user-defined firmware modules, allowing for the implementation of detector-specific protocols or leveraging pre-existing modules.

\section{Upgrade Status and Plans}

As Caribou continues to evolve, efforts are focused on enhancing its hardware, firmware, and software to meet the growing needs of the user community. This includes iterative hardware upgrades to address obsolescence and performance improvements, the development of a unified firmware framework for streamlined integration, and tools to simplify embedded software deployment. These advancements aim to reinforce Caribou’s adaptability and long-term viability across diverse detector R\&D activities.

\subsection{Hardware}
\label{subsec:hardware}
\begin{figure}[htbp]
	\centering
	\begin{subfigure}[t]{0.38\textwidth}
		\centering
		\includegraphics[width=\linewidth]{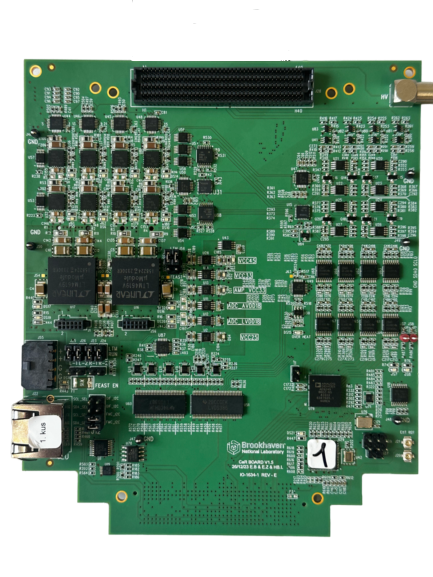}
		\caption{Picture of a CaR board v1.5}
		\label{fig:carboard_v1.5}
	\end{subfigure}
	\hfill
	\begin{subfigure}[b]{0.38\textwidth}
		\centering
		\includegraphics[width=\linewidth]{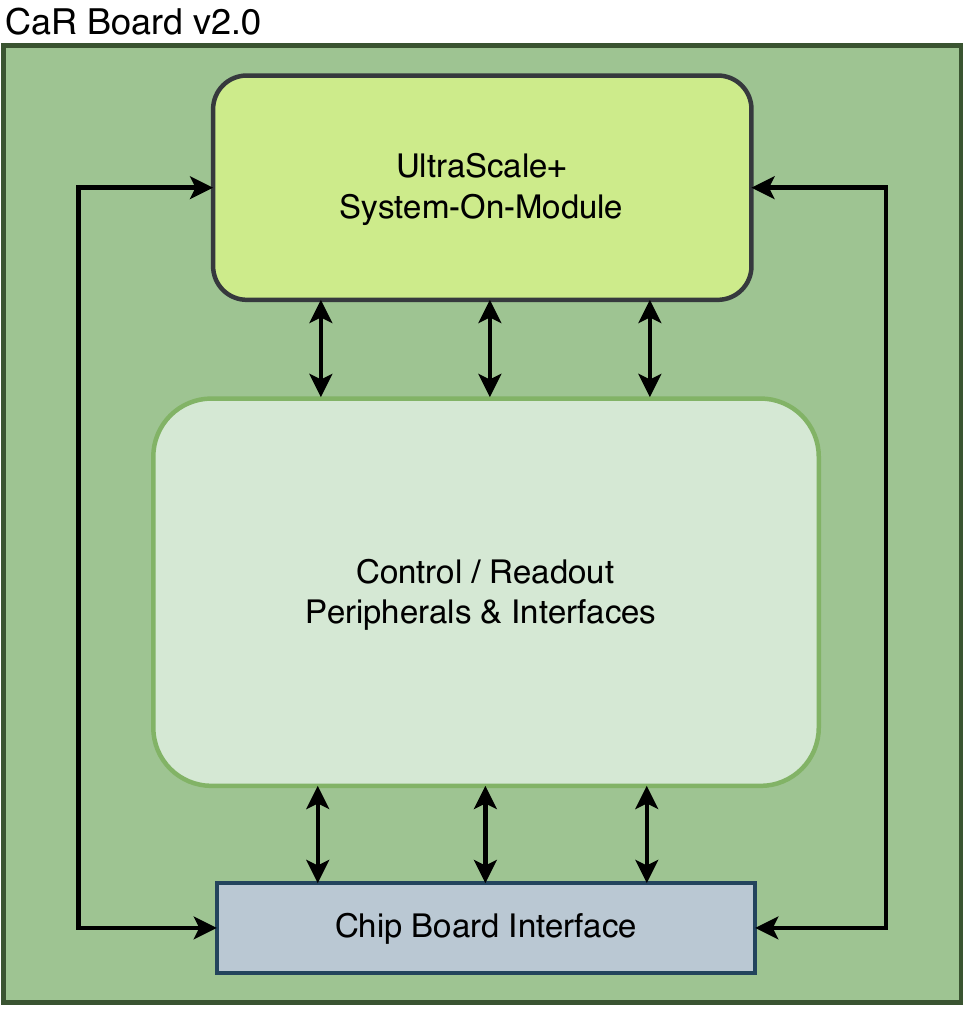}
		\caption{Simplified block diagram of the CaR board v2.0}
		\label{fig:carboard_v2.0}
	\end{subfigure}
	\caption{Mitigation of distortions and ringing affecting the bidirectional level shifters for CMOS signals}
	\label{fig:ringing_distortion}
\end{figure}

To meet the sustained demand for common hardware, an updated CaR board v1.5, shown in \mbox{Figure \ref{fig:carboard_v1.5} \cite{Carboard Gitlab}}, was designed and prototyped. The main improvements focused on replacing obsolete components and addressing signal integrity issues in the bidirectional level shifters for CMOS signals. These issues were mitigated by replacing bidirectional level shifters with unidirectional ones and adding series termination. So far, 31 boards have been produced, tested, and distributed to collaborating institutes, with an additional batch of 12 boards currently in production. 

Looking ahead, a major upgrade, Caribou v2.0, is underway to transition the system to an UltraScale+ System-On-Module (SoM) platform with a fully redesigned CaR board. The migration will occur in two phases: first, ensuring compatibility with ZCU102 and Mercury+ evaluation boards, and later integrating the SoC functionality directly into the new CaR board using SoM technology, eliminating the need for external evaluation boards. Figure \ref{fig:carboard_v2.0} shows a simplified diagram of the CaR board v2.0 design. Current efforts are focused on finalizing the board specifications and validating the layout of an initial test board integrating a subset of the circuits to be implemented in the final board. 

\subsection{FPGA Firmware}
\label{subsec:fpga_firmware}
The Boreal project is a centralized GitLab repository \cite{Boreal Gitlab} that provides a software infrastructure and a collection of useful firmware blocks to support the development and management of FPGA projects for devices using the Caribou system. Rather than enforcing a rigid firmware framework, Boreal gives users freedom to design their FPGA firmware according to their specific needs. However, it offers automation scripts for Vivado project management, simulation, synthesis and implementation, as well as a library of common cores that can be used as building blocks.

	\begin{figure}[h]
		\centering
		\includegraphics[width=1\textwidth]{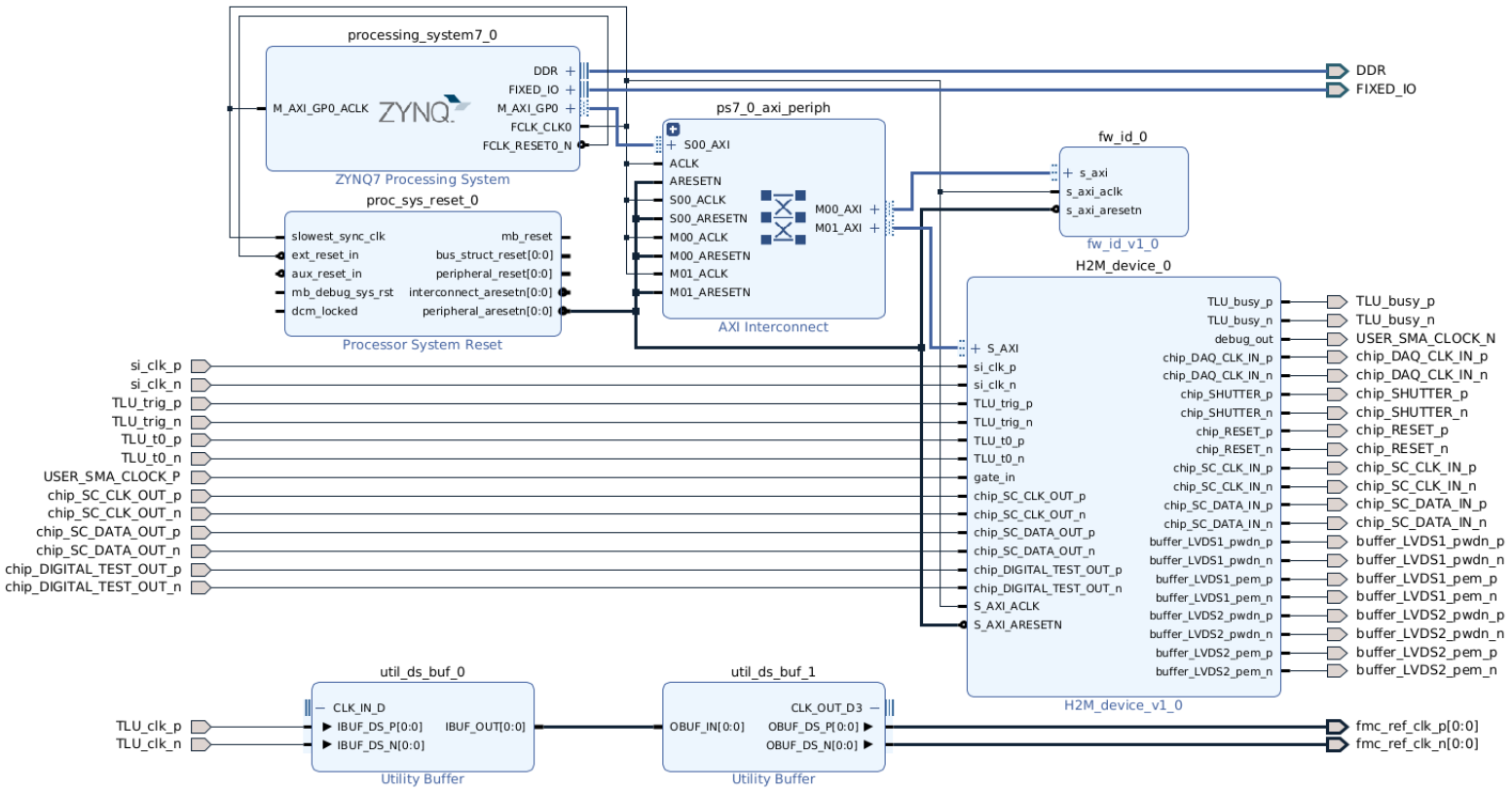} 
		\caption{Block design of the H2M \cite{Ruiz Daza} chip firmware support on the ZC706 board}
		\label{fig:h2m_block_design}
	\end{figure}
	
Each detector project is implemented as a Vivado Block Design, where the Processing System interfaces with user cores and common cores via AXI interconnects. The common cores include reusable modules and peripherals, while the user cores remains fully customizable for detector-specific logic. By providing a structured yet flexible development environment, Boreal streamlines firmware integration while allowing users to maintain full control over their designs. Figure \ref{fig:h2m_block_design} shows an example block design of the H2M \cite{Ruiz Daza} chip firmware support on the ZC706 board. 

\subsection{Software}
\label{subsec:software}
\begin{figure}[h]
	\centering
	\includegraphics[width=1\textwidth]{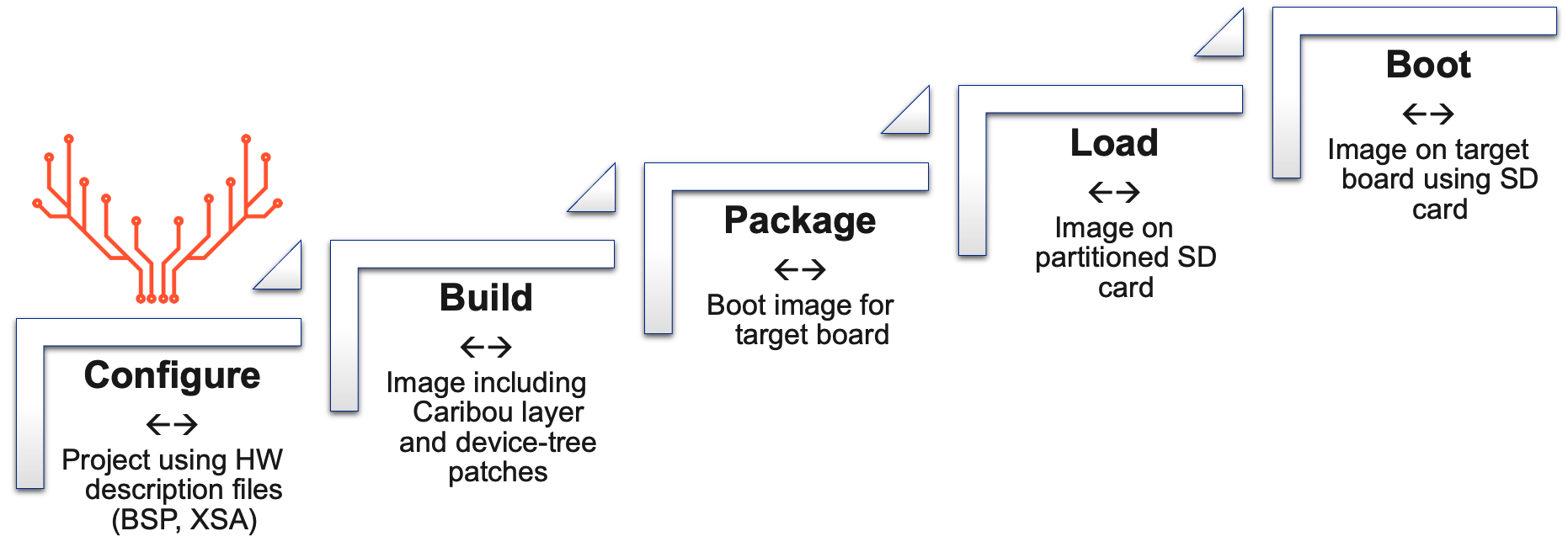} 
	\caption{Peta-Caribou build and deployment workflow}
	\label{fig:peta_caribou_worflow}
\end{figure}

In order to streamline the building of Embedded Linux images for the Caribou system, the \mbox{Peta-Caribou \cite{PetaCaribou Gitlab}} tool was designed. It relies on the AMD Xilinx PetaLinux tools and streamlines the configuration and generation of Yocto-based images for the two supported AMD Xilinx ZYNQ boards, the ZC706 and ZCU102. Peta-Caribou offers a user-friendly interface to customize and deploy system images tailored to the Caribou hardware. At the core of the project is the Peta-Manager tool, an automation script that provides a comprehensive command-line interface for configuring, building, and deploying Petalinux images. This tool ensures an efficient and seamless workflow for developers working with the Caribou system. Figure \ref{fig:peta_caribou_worflow} shows the five main steps of the Peta-Caribou Linux image build workflow.

\section{Conclusion}
In this paper, we presented the Caribou system, a modular DAQ platform for silicon pixel detector prototyping, highlighting its unified hardware, firmware, and software stack. Significant progress includes the rollout of the upgraded CaR board v1.5, the introduction of the Boreal firmware infrastructure, and the development of the Peta-Caribou image builder. Ongoing efforts focus on defining specifications and designing Caribou v2.0, which aims to transition to the UltraScale+ platform with a fully integrated SoC solution, further enhancing scalability and adaptability for future detector technologies. This project has received funding from the European Union’s Horizon 2020 Research and Innovation programme under GA no 101004761.


%
%
%
%


\end{document}